\def\ltsim{\lower.5ex\hbox{$\; \buildrel < \over \sim \;$}}
\def\gtsim{\lower.5ex\hbox{$\; \buildrel > \over \sim \;$}}
\def\ltsim{\lower.5ex\hbox{$\; \buildrel < \over \sim \;$}}
\def\gtsim{\lower.5ex\hbox{$\; \buildrel > \over \sim \;$}}
\def\kpc{$\ {h^{-1}\rm kpc}$}
\def\dndz{d{\cal N}/dz}
\def\dndm{{ $ {\rm d}^2{\rm N} / {\rm d}m{\rm d}\Omega $ }}

\documentstyle[12pt,aasms]{article}
\def\$${$$}

\def\j3s{$J_3(s)$}
\def\j3r{$J_3(r)$}

\def\etal{{\it et al.~\/}}
\def\kms{\ifmmode {\rm \ km \ s^{-1}}\else $\rm km \ s^{-1}$\fi}
\def\mpc{$\ {h^{-1}\rm Mpc}$}

\def\ltsima{$\; \buildrel < \over \sim \;$}
\def\simlt{\lower.5ex\hbox{\ltsima}}
\def\gtsima{$\; \buildrel > \over \sim \;$}
\def\simgt{\lower.5ex\hbox{\gtsima}}

\def\ltsim{\lower.5ex\hbox{$\; \buildrel < \over \sim \;$}}
\def\gtsim{\lower.5ex\hbox{$\; \buildrel > \over \sim \;$}}
\def\simprop{\lower.5ex\hbox{$\; \buildrel \propto \over \sim \;$}}

\begin{document}

\title{GALAXY CLUSTERING AROUND NEARBY LUMINOUS
QUASARS
\footnote{Based on observations with the NASA/ESA Hubble Space
Telescope, 
obtained at the Space Telescope Science Institute, 
which is operated by the Association of Universities for 
Research in Astronomy, Inc., under NASA
contract NAS5-26555.\smallskip} }

\author{Karl B. Fisher, John N. Bahcall, Sofia Kirhakos}
\affil{Institute for Advanced Study, School of Natural Sciences,
Princeton, NJ~08540}
\centerline{and}
\author{Donald P. Schneider}
\affil{Department of Astronomy and Astrophysics, The Pennsylvania 
State University, University Park, PA 16802}
\begin{abstract}
We examine the clustering of galaxies around a sample of
20 luminous low redshift ($z$\simlt $0.30$) quasars observed
with the Wide Field Camera-2 on the Hubble Space Telescope. 
The HST resolution makes possible 
galaxy identification brighter than
$V=23.5$ and as close as 2$''$ to the quasar.
We find a significant enhancement of galaxies within
a projected separation of \simlt 100~\kpc\ of
the quasars.  
If we model the qso/galaxy correlation function
as a power law with a slope given by the galaxy/galaxy
correlation function, we find that the ratio of
the qso/galaxy to galaxy/galaxy correlation functions 
is $3.8\pm 0.8$.
The galaxy counts within $r<15$~\kpc\ of the quasars
are too high for the
density profile to have an appreciable core radius (\simgt
100~\kpc).  Our results reinforce the idea
that low redshift quasars are located preferentially
in groups of 10--20 galaxies rather than in rich clusters.
We see no significant difference in the clustering amplitudes
derived from radio-loud and radio-quiet subsamples.

\end{abstract}

\vfil\eject
\section{INTRODUCTION}
\label{introduction}

Over the last two decades, it has been well established
that quasars are associated with enhancements in the
galaxy distribution.  Historically, this provided the
first direct observational 
evidence that quasars were indeed cosmological in
origin (Bahcall, Schmidt, \& Gunn 1969; Bahcall \& Bahcall 1970;
Gunn 1971; Stockton 1978) .
Over the years, considerable evidence has accumulated that 
low redshift, $z$\simlt $0.4$, quasars reside in small to moderate
groups of galaxies rather than in rich clusters
(cf. Hartwick \& Schade 1990,  Bahcall \& Chokshi 1991,
and references therein). 
At higher redshifts, there is a
marked difference in the environments of radio-loud
and radio-quiet quasars (Yee \& Green 1987). At redshifts
$z$\simgt $0.6$, radio-loud quasars are often found in
association with rich clusters (Abell richness $R\ge 1$) while
radio-quiet quasars appear to remain in smaller groups, or 
perhaps in the outer regions of clusters (Boyle \etal 1988;
Yee 1990). 

The galaxy environment around quasars provides many important
clues as to what triggers and sustains their central engines.
As first suggested by 
Toomre \& Toomre (1972), mergers and interactions
of galaxies
can provide an efficient mechanism for transporting gas
into the inner regions of a galaxy or quasar. There have
been attempts to model the interaction/merger rates
of ordinary galaxies
in order to explain the luminosity function of quasars
(De Robertis 1985; Roos 1985; Carlberg 1990) and 
the rapid evolution of the
merger/interaction rate in clusters with redshift may
provide a natural explanation of the strong evolution of
clustering observed around radio-loud quasars 
(Stocke \& Perrenod 1981; Roos 1981).  
Knowledge of the galaxy environment around
quasars is also important for understanding the
large scale distribution of quasars and how it relates to
the structure seen in galaxy surveys (Bahcall \& Chokshi 1991).

In this Letter, we examine the galaxy environment around
twenty nearby ($z$\simlt $0.3$) bright quasars observed
with the Wide Field and Planetary Camera-2 (WFPC2) of the 
Hubble Space Telescope (HST).
These fields were imaged as part of an ongoing investigation
into the nature of the host environment of quasars
(Bahcall, Kirhakos, \& Schneider 1994, 1995a, 1995b, 1996a).
The exceptional resolution of the HST images allows
companion galaxies to be detected at very close projected separations,
in some cases $r\sim 3$~\kpc\ (\simlt $2''$), and galaxy/star separation
to be performed down to $V\sim 23.5$.
The goal of the work presented here 
is to quantify the excess of galaxies associated
with these quasars.  The outline of the Letter is as follows.
A brief description of the
quasar sample used in our analysis is given in \S~2. 
In \S~3.1 we argue that galaxy counts are inconsistent
with being drawn from a uniform background.
In \S~3.2, we strengthen this conclusion by correcting
the counts for the background contamination.
We also present the excess galaxy counts above the
background in annuli of projected separation.
From these counts, we quantify the amplitude of the galaxy clustering
around the quasars in \S~3.3 in
terms of a  quasar-galaxy spatial cross correlation amplitude.
We discuss our results and their relation to previous work in
\S~4.

\section{DATA}
\label{sec-data}

The sample of objects  analyzed in this Letter
consists of 20 of the intrinsically most luminous
($M_V < -22.9$, for $H_0 = 100 {\rm kms}^{-1}{\rm Mpc}^{-1}, \
\Omega_0 = 1$) nearby ($z < 0.30$) radio-quiet and radio-loud
quasars selected from the V\'eron-Cetty \& V\'eron (1993) catalog.
Table~1 lists the individual quasars and their redshifts.
The quasars span the redshift range $0.086\le z \le 0.29$ with
a median redshift of $z_{med}=0.18$.
Details of the observations and the resulting images for eleven
of the twenty fields have been presented in 
Bahcall, Kirhakos, \& Schneider (1994, 1995a, 1995b);
the remaining observations will be presented in Bahcall \etal (1996b,
in preparation). 
Six of the quasars (denoted with an asterisk in Table~1)
are radio-loud, while the remainder are radio-quiet (Kellermann 1989).

Each quasar field was imaged with the WFPC2 through
the F606W filter, which is similar, but slightly redder than the
$V$ bandpass ($\lambda=5940$\AA, $\Delta\lambda=1500$\AA). 
The quasars were positioned within  $4''\pm 1.2''$ of the center of Wide
Field Camera CCD 3 (WF3).
Simultaneous images were obtained in the adjacent CCD chips
2 and 4 (WF2 and WF4 respectively) which together with WF3 
form a ``L'' shaped image (see figure~1). 
Each chip has $800\times 800$ pixels  
and an image scale of $0.0996''$ pixel$^{-1}$
at the chip's center; this corresponds to spatial
resolution of $2.1$~\kpc $/$arcsec at a redshift $z=0.20$ ($\Omega_0=1$).
The effective areas (areas not
covered by pyramid shadows) of WF2, WF3, and WF4 are 1.59, 1.60,
and 1.59 sq arcmin respectively. 
More detailed information on the WFPC2 and its
photometric system can
be found in Burrows (1994), Trauger \etal (1994), and Holtzman \etal (1995a,b).
The relatively long exposures (1100 or 1400 seconds), 
combined with the excellent spatial
resolution, allowed galaxies to be identified in the images down
to limiting magnitude $m(F606W) \le 24.5$ and close as  \simlt $2''$ from the 
central quasar. We performed aperture photometry on the field galaxies;
circular apertures with radii of $0.3''$ to $10''$ were
used as appropriate. 

\section{GALAXY COUNTS AROUND QUASARS}
\label{sec-counts}

\subsection{Raw Counts: Evidence for a Strong Enhancement}

If galaxies are distributed around low redshift
quasars with a power law distribution, $\xi(r)\sim
(r/10$\mpc$)^{-1.77}$ (as suggested by, e.g., Yee \& Green 1987), then
there will be a very strong enhancement of
the counts within $30''$ ($r\sim 60$~\kpc\ at $z=0.2$) of the quasar. 
Moreover, because the centers of WF2 and WF4
are offset from the quasar, there will be an enhancement
of galaxies in WF3 relative to WF2 and WF4.
The background galaxy counts rise steeply with 
magnitude and this will dilute any signal of excess galaxies.
Much of the background contamination can be removed 
simply by counting  only those galaxies with 
apparent magnitudes in the range that is likely to be physically 
associated with the quasars. 
A good compromise between eliminating too many associated galaxies and
minimizing the effect of foreground/background interlopers,
is to count galaxies in each field, $i$, which  
are in the magnitude range $m_\ast(z_i)-1$ to $m_\ast(z_i)+2$,
where $m_\ast(z_i)$ is the apparent magnitude of an $L_\ast$ 
galaxy at the redshift of the quasar, $z_i$. 
The mean (median) value of $m_\ast(z_i)$ 
for our sample
is 18.2 (18.1) 
; the total range of $m_\ast(z_i)$ is
$16.4 \le m_\ast(z_i) \le 19.2$.
In computing $m_\ast(z)$, we have taken an $L_\ast$ galaxy to correspond to 
an absolute magnitude $M_\star(F606W)=-20.75$ and have used the
K-corrections between absolute and apparent magnitudes given
in Fukugita, Shimasaku, \& Ichikawa (1995).

Figure~1 shows the positions of all galaxies in our twenty fields
with $m_\ast(z_i)-1\le m \le m_\ast(z_i)+2$;
geometric corrections were applied according to Holtzman \etal (1995a).
In the panel containing the quasar (WF3, lower left), 
there is a clear excess of galaxies (50 galaxies) relative to 
WF2 (upper left) and WF4 (18 and 17 galaxies respectively). 
The hypothesis that the counts in all three chips 
are drawn from an underlying Poisson distribution with a 
common mean leads
to a maximum likelihood mean per chip of 28.4 and a chi-squared of
$\chi^2=24.5$; the probability that $\chi^2$ for
one degree of freedom is this large by chance is only
$P=7\times 10^{-7}$.  Thus, without any detailed 
modeling, we can rule out the possibility that the  counts are
random fluctuations in the background distribution. In the following
sections, we attempt to make this conclusion progressively more 
quantitative by first modeling the background galaxy distribution and
then introducing a model for the spatial distribution 
of galaxies around the quasars.

\subsection{Correcting for the Background Galaxies}

In order to further quantify the excess of galaxy counts
around the quasars, we need an estimate
of the contribution from background galaxies.  
In figure~2, we show the galaxy counts (\dndm, per
square arcsec) versus magnitude for the
``off quasar'' chips WF2 and WF4, in bins $\Delta m(F606W) =0.5$. 
The counts at $m(F606W) >21.5$ mag 
are well approximated by a power law, 
$\log_{10}$\dndm $= -10.8 + 0.33\, m(F606W)$; 
the counts at brighter magnitudes are
in excess of those obtained by  extrapolating the faint power law fit. 
Because the separation of the centers of chips 2 and 4 is
not large ($101''$ corresponds to 213~\kpc\ at $z=0.2$), 
the counts in these chips are a combination of both background
galaxies and the (relatively bright) 
galaxies physically associated with the quasars, and
hence yield an overestimate of the true background. 
We have compared the power-law fit 
with the counts derived from the HST 
Medium Deep Survey (MDS) (Griffiths \etal 1994; S. Casertano,
private communication) which covers a much larger area of the sky.
The agreement is good, and in the remainder of this Letter, we will adopt the
power law in figure~2 as our estimate of \dndm. The agreement
with the MDS is also a useful consistency check for systematic
errors in our derived magnitudes. 

In the twenty fields, there
are 11 galaxies in the range $m_\ast(z_i)-1 \le m \le m_\ast(z_i)+2$ 
within a projected separation less than 25~\kpc\ of
the quasar; the total number expected  
from the \dndm\ power law relation is only $0.99$. The
probability of the observed counts being a Poisson realization
of the background is extremely small, $P=8\times 10^{-9}$, 
two order of magnitudes smaller than our
previous estimate obtained by neglecting the
background. This is also a much stronger constraint 
than the upper limit given in Bahcall, Kirhakos,
\& Schneider (1995a), $P=2\times 10^{-2}$,   
based on eight fields and an (over)estimate
of the background obtained from the counts in
WF2 and WF4.

We have also computed counts within projected
separations of 5 and 10~\kpc. In the twenty
fields, we find $2/0.039$ and $5/.16$ (number/number expected in the
background) with $r<5$~\kpc\ and $r<10$~\kpc\ respectively; the
corresponding random probabilities 
are $P=7\times 10^{-4}$ and $P=7\times 10^{-7}$.
There is a suggestive difference in the
counts derived from the radio-loud and radio-quiet
subsamples. The radio-quiet quasars (14 fields) had
$0/.027$ ($P=0.97$) and $2/.11$ ($P=5\times 10^{-3}$) for the
5 and 10~\kpc\ counts; the corresponding numbers for
the radio-loud sample (6 fields) were
$2/.012$ ($P=8\times 10^{-5}$) and $3/.05$ ($P=2\times 10^{-5}$).

In figure~3, we show the fractional excess of galaxies above
the background, $\langle \delta N/N(r) \rangle = 
\langle N_i(r)/N_{b,i}(r)\rangle-1$, obtained by
averaging the counts in the twenty fields 
in bins of 15~\kpc\ projected separation.
Here $N_i(r)$ is the
actual count of galaxies with projected separations
between $r-\Delta r/2$ and $r+\Delta r/2$
in the i$^{\rm th}$ field and $N_{b,i}(r)$ 
is the expected background contribution.
Again, we have only considered those galaxies in the apparent
magnitude range $m_\ast(z_i)-1$ to $m_\ast(z_i) +2$. 

From figure~3, we can immediately draw two conclusions.
First, there is a significant excess of galaxies 
within projected separations of $r<$\kpc\ from the quasars.
Second, there appears to be no significant difference
in the galaxy counts for the radio-loud and radio-quiet
subsamples for 10~\kpc\ $<r< 100$~\kpc. 
In the next section, we quantify the
observed clustering in terms of the spatial quasar/galaxy
cross-correlation function.

\subsection{Estimating the Spatial Clustering Amplitude}
\label{sec-xi}

A detailed derivation
of the relation between angular counts
and a spatial distribution of galaxies in terms
of a cross-correlation function is given in Longair \&
Seldner (1979). Briefly, one assumes that the galaxy distribution 
(above the background) around the quasar is specified     
by a quasar/galaxy cross-correlation function, $\xi_{qg}(r,z)$. 
The observed excess in projected separation in the $i^{\rm th}$
field is then
obtained by integrating $\xi_{qg}(r,z)$ over the 
redshift distribution of galaxies $\left(\dndz\right)_i$ in
the apparent magnitude range 
$[m_\ast(z_i)-1\le m \le\ m_\ast(z_i)+2]$, 
\begin{equation}
 N_i(r)/N_{b,i}(r) - 1 = 
{{\int d\theta dz\,  \left(\dndz\right)_i
\, 
\xi_{qg}(s,z)}\over{
\int d\theta dz\,  \left(\dndz\right)_i }} \qquad .
\label{eq-deltan}
\end{equation}
In this equation, $s$ is the comoving separation
between between the line of sight, $x(z)$, and the
quasar located at coordinate $x(z_i)$, 
$s^2 = \left\{(x(z_i)-x(z))^2/G[x(z)]\right\} + 
x(z)^2\theta^2$; $\theta$ is the angle between
the line of sight and the quasar and $G(x)$ is a function which
describes the degree of spatial curvature
$G(x)=1-(H_0x/c)^2(\Omega_0-1)$ (cf. Peebles 1980,
eqn. 56.13).  The range of integration in $\theta$ corresponds to
the angles which span the bin of projected
separation, i.e., $(r-\Delta r/2)/d_a(z_i)$ to $(r+\Delta r/2)/d_a(z_i)$
where $d_a(z_i)$ is the angular diameter distance
to the quasar.

The redshift distribution $\left(\dndz \right)_i$
(assuming no galaxies are created or destroyed)
is given by 
\begin{equation} 
\left( \dndz\right)_i 
= \phi(z_i)\, {{dx}\over{dz}}\, dV
\qquad ,
\end{equation}
where $\phi(z_i)$ is the integral of the luminosity
function over the absolute magnitudes 
$[M_\ast(z_i)-1\le M\le\ M_\ast(z_i)+2]$
corresponding the
the apparent magnitude range $[m_\ast(z_i)-1\le m\le\ m_\ast(z_i)+2]$,
\begin{equation}
\phi(z_i) \equiv
\int_{M_\ast(z_i)-1}^{M_\ast(z_i)+2} dM\, \Phi(M) \qquad .
\end{equation}

The cosmological model $(\Omega_0, \Lambda, H_0)$ enters in 
the above equations implicitly 
in the volume element, angular distance, and luminosity distance.
We adopt a canonical model with 
$\Omega_0=1,\ H_0 = 100\ {\rm kms}^{-1}{\rm Mpc}^{-1}$, and $\Lambda=0$.
The basic results of our study are largely independent of
this choice.
The redshift distribution is computed using a 
non-evolving Schechter (1976) luminosity function
with a faint end slope of $\alpha = 0.97$ (Loveday \etal 1992).
The K-corrections necessary for computing the
relation between absolute and apparent magnitudes have been
taken from Fukugita, Shimasaku, \& Ichikawa (1995).
In the results that follow, the $H_0$ dependence is explicitly indicated.
The derived clustering amplitudes remain within the stated
errors as $\Omega_0$ is varied from $0.1$ to $1.0$. 

The cross-correlation function, $\xi_{qg}(r,z)$, is
usually assumed to evolve with redshift.  
A convenient model for this evolution is to assume 
that on small scales the clustering is constant in
physical coordinates, i.e. the number of excess galaxies
around the quasar, $n(z)\xi_{qg}(r,z)$, is a
constant ($n(z)$ is average number density at the epoch $z$). 
Thus for any assumed shape of the correlation function,
$F(r)$, $\xi_{qg}(r,z)$ evolves as
\begin{equation}
\xi_{qg}(r,z) =  {{1}\over{(1+z)^3}}\ F\left[ {{r}\over{1+z}} \right]
\qquad .
\end{equation}
The factor of $(1+z)^3$ compensates for the change in
the mean number density, while
the redshift factor in the argument of $F(r)$ is simply a 
matter of convention; one usually specifies the shape
in terms of {\it physical} separations while in 
(\ref{eq-deltan}), we have specified $\xi_{qg}(r,z)$ in terms of
comoving separation. 

We consider three different models for 
$F(r)$. The first model is a power law, $F(r) = B_{qg}r^{-1.77}$,
with a slope equal to that of the galaxy/galaxy correlation
function (Davis \& Peebles 1983) and the amplitude taken as
a free parameter. The second model is
an exponential surface density of galaxies, $\mu(r) = \mu_0 
\exp(-r/r_c)$ which, after deprojection by a standard Abel
inversion, corresponds to $F(r)= \mu_0/\pi r_c K_0(r/r_c)$
($K_0(x)$ being the modified Bessel function).
This model was proposed by Merrifield \& Kent (1989) as
a typical galaxy profile around centrally dominant cluster
galaxies. We adopt their best estimate of a core radius
of $r_c=100$~\kpc, and vary the amplitude $\mu_0$.  Last, we
consider a modified Hubble profile $F(r) = A\, [1 +(r/r_c)^2]^{-3/2}$
(Binney \& Tremaine 1987, eqn. 2-37) with $r_c=100$~\kpc\ and $A$
a free parameter.  These three models have different behavior at small $r$.
Both the power law and exponential models diverge 
as $r\to 0$ (although the later does so only weakly, $K_0(x) \sim -\ln(x)$),
while the modified Hubble model asymptotically approaches a constant.

We varied the amplitude of the models to achieve a maximum
likelihood fit to the excess galaxy counts in bins of
projected separation of 15~\kpc. The limited number of
fields prevented us from varying more than one parameter per model.
Figure~3 shows the resulting best-fit models; the amplitudes of
the power-law model are given in Table~2. At separations
$r$\simgt 30~\kpc, the excess counts are relatively flat and
all three models for the correlation function
fit the data. However, the counts in the
innermost bin $r<15$~\kpc\ lie significantly above the
counts at larger separations; the rise in the counts at
small radii is particularly striking in the five radio
loud fields. The power-law is the only
model for $\xi_{qg}(r)$ considered which rises steeply enough
to account for this excess. 

\section{DISCUSSION}
\label{sec-disc}

If the quasars were distributed like typical galaxies, then
the derived value of $\langle B_{qg}\rangle$ would be equal to the amplitude
of the galaxy/galaxy correlation function, $\langle B_{gg}\rangle\sim 20$
(Davis \& Peebles 1983). A higher value of $\langle B_{qg}\rangle$
suggests that the quasars lie preferentially in regions of
above average galaxy density.  
Following Bahcall \& Chokshi (1991),  
we can convert our values of $B_{qg}$ into an estimate
of the  typical richness of quasar galaxy environment. 
Here, we define richness
as the number of $L_\ast$ galaxies associated with the 
quasar; this is given by a  
simple integral over the correlation function,
$N= 4\pi n_\ast \int \xi_{qg}(r)r^2 dr$, where the limits
of integration are $r=0$ to 1.5 \mpc\ (the traditional Abell radius), and 
$n_\ast \approx 1.5\times 10^{-2}{\rm h}^3{\rm Mpc}^{-3}$ is the
number density of $L_\ast$ galaxies. Using the values of
$\langle B_{qg}\rangle $ in Table~2, we find the quasars typically reside in 
groups of 16--25 galaxies. These numbers should
be compared to the typical richness of Abell clusters which
have 30--49 and 50--79 members for richness classes $R=0$ and
$R=1$ respectively.  Moreover, this estimate is most likely
an upper limit since there is evidence that the galaxy profile around quasars 
falls off more steeply than $r^{-1.77}$ for $r$\simgt 0.25 \mpc\
(Ellingson, Yee, \& Green 1991). In order to test the robustness
of our results to a steepening of the galaxy profile, we 
fitted a double power law model with slope of 
${-1.77}$ for $r\le 0.25$ \mpc\ and $-3$ for $r>0.25$ \mpc.
The best fit amplitudes, $B_{qg}$, for this model increased by about 10\%
(still well within the quoted $1-\sigma$ errors), yet, the
number of infered bright galaxies associated with the quasars decreased
to 8 due to the steeper profile at large separations.

The derived amplitudes, $\langle B_{qg}\rangle $, in the pure power law case 
are somewhat larger than the estimates by Yee \& Green (1987). They examined
the clustering of 9 radio-loud and 16 radio-quiet quasars in the redshift 
range $0.15<z<0.30$. 
On scales of $\sim$ 20--500~\kpc, they estimated 
$\langle B_{qg}\rangle \approx 60\pm 20$ and $\langle B_{qg}\rangle\approx 42\pm 14$ for radio-loud and
radio-quiet quasars respectively.  
Hayman (1990) derived galaxy
counts around low redshift ($z<0.3$) quasars from the Palomar
Sky Survey prints.  He found that the ratio of the quasar/galaxy and
galaxy/galaxy angular correlation functions was $3.1\pm 0.6$;
if the quasars and galaxies have similar selection functions, this
translates to an estimate
similar to Yee \& Green of  $\langle B_{qg}\rangle\sim 61\pm 12$. 
French \& Gunn (1983) analysed a sample 25 low-redshift quasars 
($z\simlt 0.35$) selected from 1.2-m Palomar Schmidt plates
and concluded $\langle B_{qg}\rangle= 25 \pm 12$; they also
analysed the data set of Stockton (1978, 27 quasars 
with redshifts $z\le0.45$ selected from the red Sky Survey
prints) and 
derived, via the same analysis, $B_{qg}=79\pm 40$. 
Our measurements are, with the exception of Yee \& Green's
value for radio-quiet quasars, consistent within the quoted
errors.  The slightly higher clustering amplitude we derive 
for the radio-quiet subsample may be a result of our 
sample being the subset of the most luminous quasars.

Yee \& Green (1987) found that the clustering amplitude of
galaxies around radio-loud quasars increased by a factor $\sim 3$
between $z\sim 0.4$--$0.6$ and at $z\sim 0.6$ radio-loud
quasars are found in 
environments as rich as Abell class $R=1$. Optical quasars
do not evolve as rapidly (Boyle \etal\ 1988), perhaps
indicating a different formation scenario.
It has been suggested the quasars and active galactic
nuclei may be triggered by interactions 
(e.g., Toomre \& Toomre 1972;
Stocke \& Perrenod 1981; Roos 1981, Yee 1987).  
This offers a simple explanation for why the quasars are 
typically not found in  
rich clusters at low redshifts; the high
velocity dispersion of such clusters leads to a low  
interaction rate. 

The HST WFPC2 is an excellent instrument for extending the present
analysis to fainter, low-redshift, quasars. This extension would
improve the counting statistics while also
providing information regarding possible correlations of the
quasar environment with luminosity. The imaging of
of moderate redshift quasars ($z\simlt 0.6$) could be
accomplished by HST with single orbit exposures. This imaging
would provide a more direct comparison with previous
ground based work and would increase our knowledge
of the evolutionary history of the quasar environment.
An increased 
knowledge of the quasar environment would be useful
in using quasar observations to probe large
scale structure at higher redshifts.

\acknowledgments
We would especially like to thank Stefano Casertano and
the members of the Medium Deep Survey for useful
comments and for allowing us to compare our
number counts and magnitudes with their measurements.
We have also benefited from discussions with
Neta Bahcall, Michael Strauss, Ofer Lahav. We thank
the anonymous referee for helpful suggestions.
This work was supported in part by NASA contract 
NAG5-1618, NASA grant number NAGW-4452,
and grant number GO-5343 from the Space Telescope Science
Institute, which is operated by the Association of Universities for
Research in
Astronomy, Incorporated, under NASA contract NAS5-26555.
KBF acknowledges the support of the NSF.

\vfil\eject

\vfill\eject

\begin{table*}[tbh]
\centering
\begin{tabular}{|lcccc|}
\multicolumn{5}{c}{{\bf TABLE 1}}\\
\multicolumn{5}{c}{{\sc quasar sample}}\\ 
\multicolumn{5}{c}{{ }}\\ \hline 
Name                  &   Date    & Exp. (sec) &\ \ \ \  $M_V$ \ \ \  \
&\ \ \ \ \ \   z \ \ \ \ \ \  \\  \hline
PG 0052+251           & 05 DEC 94 &   1400     & $-23.0$ & 0.155\\
PHL 909               & 17 OCT 94 &   1400     & $-22.9$ & 0.171\\
NAB 0205+02           & 26 OCT 94 &   1400     & $-23.0$ & 0.155\\
0316$-$346$^\ast$     & 20 NOV 94 &   1400     & $-24.5$ & 0.265\\
PG 0923+201           & 23 MAR 95 &   1400     & $-23.0$ & 0.190\\ 
PG 0953+414           & 03 FEB 94 &   1100     & $-24.1$ & 0.239\\
PKS 1004+13$^\ast$    & 26 FEB 95 &   1400     & $-24.3$ & 0.240\\
PG 1012+00            & 25 FEB 95 &   1400     & $-23.0$ & 0.185\\
HE 1029$-$1401        & 06 FEB 95 &   1400     & $-23.2$ & 0.086\\
PG 1116+215           & 08 FEB 94 &   1100     & $-23.7$ & 0.177\\
PG 1202+281           & 08 FEB 94 &   1100     & $-23.0$ & 0.165\\
3C 273$^\ast$         & 05 JUN 94 &   1100     & $-25.7$ & 0.158\\
PKS 1302$-$102$^\ast$ & 09 JUN 94 &   1100     & $-24.6$ & 0.286\\
PG 1307+085           & 05 APR 94 &   1400     & $-23.1$ & 0.155\\
3C 323.1$^\ast$       & 09 JUN 94 &   1100     & $-22.9$ & 0.266\\
PG 1309+355           & 26 MAR 95 &   1400     & $-23.2$ & 0.184\\
PG 1402+261           & 07 MAR 95 &   1400     & $-23.0$ & 0.164\\
PG 1444+407           & 27 JUN 94 &   1100     & $-24.0$ & 0.267\\
PKS 2135$-$147$^\ast$ & 15 AUG 94 &   1400     & $-23.5$ & 0.200\\
PKS 2349$-$014$^\ast$ & 18 SEP 94 &   1400     & $-23.3$ & 0.173\\ \hline\hline
\multicolumn{5}{l}{$^\ast$ Radio loud}
\end{tabular}
\end{table*}
\vfill\eject

\begin{table*}[tbh]
\centering
\begin{tabular}{lccc}
\multicolumn{4}{c}{{\bf TABLE 2}} \\
\multicolumn{4}{c}{{ \sc Clustering Amplitudes  }}\\
\multicolumn{4}{c}{                              }\\  \hline\hline 
Sample      & $\langle B_{qg}\rangle^\dagger$ 
& $1-\sigma$ Range & $
\langle B_{qg}\rangle/\langle B_{gg}\rangle^{\dagger\dagger}$ \\ \hline
All quasars &   75          &   60--93            & $3.8\pm 0.8$      \\
Radio Loud  &   84          &   57--117           & $4.2\pm 1.5$     \\ 
Radio Quiet &   72          &   53--92            & $3.6\pm 1.0$\\ \hline
\multicolumn{4}{l}{$^\dagger$in units of [\mpc]$^{1.77}$  }\\
\multicolumn{4}{l}{$^{\dagger\dagger} \langle B_{gg}\rangle
=19.8 $(\mpc)$^{1.77}$ }\\
\end{tabular}
\end{table*}

\clearpage
\centerline{FIGURE CAPTIONS}

Fig. 1.--- Mosaic of the galaxy counts in WF2, WF3, and WF4
(upper left, lower left, and right panels respectively) for all
twenty fields. In a given field $i$, we count only those galaxies in the
magnitude range $m_\ast(z_i)-1$ to $m_\ast(z_i)+2$ where $m_\ast(z_i)$
is the apparent magnitude of an $L_\ast$ galaxy at the redshift
of the quasar, $z_i$.

Fig. 2.--- Number counts, \dndm,  versus $m(F606W)$. 
The filled circles represent the counts determined directly from the off chips in each QSO field.  Poisson errors are shown. The line is a least squares 
fit to the faint counts with $m(F606W)>21.5$.

Fig. 3.--- Average galaxy counts $\langle\delta N/N(r)\rangle$ of the
full quasar sample and radio and radio-quiet subsets.
The solid points are the counts computed using a power
law fit to the faint counts of chips 2 and
4 for the background (cf. figure~2); the open symbols (shifted .025 in the log) 
show the counts derived using the background counts from
the Medium Deep Survey.
The error bars show the scatter between the different fields.
The curves are the predicted correlation functions
if the galaxy/quasar cross correlation function is 
a power law with index $\gamma=1.77$ (solid), 
an exponential surface density (dashed), or a modified Hubble profile
(dotted).  All calculations assume $\Omega_0=1$ and $\Lambda=0$.

\end{document}